\documentclass{article}

\usepackage{amsmath}
\usepackage{amsfonts}
\usepackage{graphicx}

\usepackage{caption}
\usepackage{subcaption}

\usepackage{authblk}
\usepackage{amssymb}

\usepackage{epstopdf}
\usepackage{epsfig}
\usepackage{hyperref}

\usepackage{color}

\date{\today}

\begin{document}

\title{Axial Quasi-Normal Modes of Scalarized Neutron Stars with \\
Realistic Equations of State}

\author{Zahra Altaha Motahar}
\author{Jose Luis Bl\'azquez-Salcedo}
\author{Burkhard Kleihaus}
\author{Jutta Kunz}

\vspace{0.5truecm}
\affil{Institut f\"ur  Physik, Universit\"at Oldenburg, Postfach 2503,
D-26111 Oldenburg, Germany}

\vspace{0.5truecm}

\vspace{0.5truecm}

\vspace{0.5truecm}
\date{
\today}

\maketitle

\begin{abstract}

We compute the axial quasi-normal modes of static neutron stars 
in scalar tensor theory. In particular, we employ various realistic equations 
of state including nuclear, hyperonic and hybrid matter. 
We investigate the fundamental curvature mode and compare the results 
with those of General Relativity. We find that the frequency of the modes 
and the damping time are reduced for the scalarized neutron stars. 
In addition, we confirm and extend the universal relations for 
quasi-normal modes known in General Relativity 
to this wide range of realistic equations of state for 
scalarized neutron stars and confirm the universality of the 
scaled frequency and damping time in terms of the scaled moment of inertia
as well as compactness 
for neutron stars with and without scalarization. 
\end{abstract}

\section{Introduction}

Neutron stars represent some of the most interesting objects in the universe. 
Their high compactness and density make these objects a valuable 
astronomical laboratory for studying gravity and testing 
General Relativity (GR) and alternative theories of gravity 
in the strong gravity regime. On the other hand, neutron stars 
consist of extremely dense matter that cannot be produced 
in ground based laboratories, resulting in our current poor 
understanding of the physics of matter at supranuclear densities.
 
Recent gravitational wave (GW) detections by the LIGO and VIRGO collaboration 
\cite{ Abbott:2016blz, Abbott:2016nmj, Abbott:2017vtc, Abbott:2017oio, 
TheLIGOScientific:2017qsa, Abbott:2017dke, Abbott:2017gyy, GBM:2017lvd} 
have provided us with an opportunity to test GR in the strong gravity regime 
as well as to constrain the equation of state (EOS) of high density matter. 
In particular, GWs from the binary neutron star merger GW170817 
\cite{TheLIGOScientific:2017qsa}, which was also observed 
by a $\gamma$-ray burst (GRB 170817A), have opened a new window 
in astrophysics, namely multi-messenger astronomy, to study the properties 
of compact objects \cite{GBM:2017lvd,Coulter:2017wya}. 
In general, there are several possibilities for the remnant 
of such a binary neutron star merger. The remnant may either 
be a black hole, an unstable massive neutron star which eventually 
collapses to form a black hole, or a stable neutron star 
\cite{Abbott:2017dke,Piro:2017zec}. 

In order to shed more light on this question, one has to detect 
the GW radiation after the merger during the ringdown phase 
and identify the specific frequencies and damping times. However, 
in order to be able to observationally detect the ringdown frequencies 
and damping times of such events
the next generation of GW detectors must be awaited.
Yet it is essential to improve our theoretical understanding 
of the ringdown phase of these compact objects and make 
theoretical predictions that can be tested by future GW detections. 

The radiation emitted in the collapse of a star and in the coalescence 
of binaries has a close relation to the quasi-normal modes (QNMs) 
of the final compact object 
\cite{Thorne:1967a,Thorne:1969a,Thorne:1969b,Thorne:1969rba,Lindblom:1983ps,Detweiler:1985zz,Chandrasekhar:1991fi,Chandrasekhar:1991a,Chandrasekhar:1991b,Ipser:1991ind,Kojima:1992ie,Kokkotas:2003mh,Kokkotas:1994an}.
QNMs are the natural oscillations of physical systems on astrophysical 
scales such as relativistic compact objects. By using the QNMs formalism, 
one can study the damped harmonic pulsations giving rise to GWs 
(see e.g.~\cite{Kokkotas:1999bd,Nollert:1999ji,Ferrari:2007dd}). 
Stellar oscillations coupled with GWs are usually damped out, 
since the waves carry away the pulsational energy of the star. 

The QNM spectrum of an astrophysical object depends only on its properties. 
In the case of a neutron star, which contains matter, it is different 
and richer than in the case of a black hole. 
The particular frequencies and damping times of the ringdown 
depend on the composition of the matter inside the star 
(see e.g.~\cite{Heiselberg:1999mq,Haensel:2007yy,Lattimer:2012nd}). 
The eigenfrequencies of the modes of the system are obtained 
as complex numbers, where the real part represents the frequency 
of the oscillation, and the inverse of the imaginary part 
gives the damping time.

In the static case the perturbations of the spacetime can be classified 
in terms of \textit{axial} (odd parity) and \textit{polar} (even parity) modes.
Whereas the polar perturbations couple to the fluid, 
the axial perturbations of the gravitational field do not couple 
to the density and pressure perturbations of the fluid.
Thus they do not excite any fluid pulsations inside the star.

The polar modes of non-rotating compact stars include the fundamental $f$-mode 
together with the pressure $p$-modes, and the gravity $g$-modes, 
which are all present already in Newtonian stars \cite{Cowling:1941a},
and in addition the spacetime $w$-modes (GW modes), which are without
analogue in Newtonian stars but can exist only in General Relativity (GR)
\cite{Kokkotas:2003mh} and its generalizations.
These $w$-modes are characterized by high frequencies and 
rapid damping of the oscillations
\cite{Kokkotas:2003mh}.

In this paper, we focus on the axial modes of neutron stars. 
The axial modes do not possess an analogue in Newtonian theory, either.
Representing a pure manifestation of the tensorial character of gravity
the standard $w$-modes are curvature modes.
As for black holes, 
the axial modes of neutron stars can be obtained from a single
wave equation with an effective potential.
Forming (presumably) an infinite tower of modes,
their eigenfrequencies depend, however, on the EOS 
\cite{Kokkotas:1999bd,Nollert:1999ji}.

But QNMs can provide us not only with valuable information 
on the composition of neutron stars. 
They can also be utilized to learn about possible modifications of GR, 
as discussed, for instance, in recent reviews on 
generalized theories of gravity with an emphasis 
on testing the strong gravity regime with GWs
\cite{Berti:2018cxi,Berti:2018vdi}.
There are numerous ways for modifying gravity 
such as adding one or more scalar fields, tensor or vector fields, 
going to higher orders, etc. (see e.g.~the reviews
\cite{Capozziello:2011et,Berti:2015itd}).
Besides in GR, axial QNMs have been studied in several alternative theories 
of gravity, such as tensor-vector-scalar theory \cite{Sotani:2009nm}, 
Einstein-Gauss-Bonnet-dilaton theory \cite{Blazquez-Salcedo:2015ets}, 
Horndeski gravity \cite{Blazquez-Salcedo:2018tyn} 
and recently $f(R)$ gravity (quadratic gravity) 
\cite{Blazquez-Salcedo:2018qyy}.

Although properties of neutron stars have mostly been studied in GR, 
it is certainly also of interest to study them in 
alternative theories of gravity \cite{Berti:2015itd}. 
For instance, neutron stars in scalar-tensor theory (STT) 
may feature besides the usual GR solutions nontrivial scalarized solutions
that arise spontaneously \cite{Brans:1961sx,Damour:1992we,Fujii:2003pa}. 
This phenomenon called spontaneous scalarization 
was found by Damour and Esposito-Farese \cite{Damour:1993hw,Damour:1996ke}.
Here under suitable conditions scalarized neutron stars will arise,
since the scalar field nonlinearities may intensify the
attractive nature of the scalar field interactions.

The physical properties of scalarized neutron stars may then deviate
from the basic neutron star properties in GR, as shown in 
\cite{Damour:1993hw,Damour:1996ke,Harada:1998ge,Salgado:1998sg,Sotani:2012eb,Pani:2014jra,Sotani:2017pfj}
for static and slowly rotating neutron stars.
Doneva et al. 
\cite{Doneva:2013qva,Doneva:2014uma,Doneva:2014faa,Staykov:2016mbt,Yazadjiev:2016pcb,Doneva:2016xmf}
have also studied rapidly rotating neutron stars in STT as well 
as scalarized neutron stars with a massive scalar field
and found that the effect of scalarization becomes more enhanced. 
As a simple and natural extension of GR, we here consider STT 
by adding only a single massless scalar field as an additional mediator 
of the gravitational interaction.

Scalar gravitational waves from relativistic stars in STT
have been studied for the first time by Sotani \cite{Sotani:2005qx}. 
The study has shown, that
if such scalar GWs would exist, they might be quite weak and hard 
to detect by the current generation of GW detectors. 
It has also shown that the spontaneous scalarization 
could even be observed for radial oscillations.
In addition, the merger of neutron stars in $R^2$ gravity has been studied recently in \cite{Sagunski:2017nzb}, where it has been shown that during the ringdown phase the radial scalar mode can even dominate over the fundamental mode.
All this indicates that, with the help of further observations of quantities 
such as the stellar mass or compactness,
it could be possible to extract the imprint of the gravitational theory from the detection of the ringdown phase of a neutron star.

However the difficulty lies in the uncertainty of the composition of the star in its core.
When constructing neutron star models,
the specific choice of the EOS employed 
leads to distinct properties of the object, including the ringdown frequencies and damping times. 
Choosing the proper EOS and testing different proposed EOSs 
is a very active field in gravity and astrophysics,
as well as in nuclear physics.
Due to our lack of knowledge about the physical properties of matter 
at very high densities, there are a large number of proposed EOSs.

Nevertheless, 
it is possible to construct several universal relations for neutron stars, 
i.e., relations among properly scaled physical quantities 
that are to a large extent independent of the EOS. Deviations of these relations in alternative theories of gravity from GR 
let us, in principle, test the gravity theory as well as infer 
more information from the observational data
(see e.g.~the recent reviews \cite{Yagi:2016bkt,Doneva:2017jop}).
Starting with the early work in 
\cite{Andersson:1996pn,Andersson:1997rn,Kokkotas:1999mn,Benhar:1998au,Benhar:2004xg,Tsui:2004qd}
a variety of universal relations for 
oscillation frequencies in GR have been found,
covering a large range of realistic EOS
\cite{BlazquezSalcedo:2012pd,Blazquez-Salcedo:2013jka}. 
These could be exploited, for instance, 
once the frequency or damping time can be extracted from detected GWs, 
to find the mass, radius or moment of inertia of compact stars
and thus infer their EOS.

In fact, there exist further universal relations, 
such as the ''I-love-Q'' relations \cite{Yagi:2013bca},
between various parameters of neutron stars,
which are known to hold approximately for most 
realistic matter models in GR
\cite{Yagi:2016bkt,Doneva:2017jop}.
However, when alternative theories of gravity are considered,
the corresponding universal relations may deviate from those of GR 
\cite{Yagi:2013bca,Will:2014kxa,Blazquez-Salcedo:2015ets,Yagi:2016bkt,Doneva:2017jop}.
Therefore making use of potential deviations in universal relations
together with further observations can, in principle, constrain 
alternative theories of gravity. 

In this paper we present the spectrum of axial QNMs 
of static and spherically symmetric scalarized neutron stars. 
In section \ref{sec:Basic Eqs} we establish 
the mathematical and physical framework for the neutron star models.
Here the STT action 
is defined in the Einstein frame, and the relation between 
the Jordan and the Einstein frame is given.
We then discuss the axial perturbations 
and the basic equations for axial QNMs 
of static neutron stars in STT,
presenting the differential equations and boundary conditions.
In the following subsections, we explain the set of realistic EOSs employed,
and briefly describe the numerical method.
We present our results 
for the scalarized neutron star models as well as for the GR solutions
in section \ref{sec:Results}.
Here we also investigate several universal relations 
involving the (scaled) axial QNMs.
In section \ref{sec:Conclusions} we represent 
a summary of our results.

\section{The model}\label{sec:Basic Eqs}

We now provide the theoretical framework for the study.
We recall the equations to obtain equilibrium neutron stars in STT, 
consider axial perturbations in STT, and discuss the various EOS
employed as well as the numerical method.

\subsection{Scalar-tensor theory}\label{sec:STT}

In this subsection we address neutron star models in STT.
In the Einstein frame the action is given by
\begin{equation}
S= \frac{1}{16\pi G}\int d^4x \sqrt{-g} \left[{\cal R} -
2g^{\mu\nu}\partial_{\mu}\varphi \partial_{\nu}\varphi\right]+ S_{m}[\Psi_{m}; \mathrm{A}^{2}(\varphi)g_{\mu\nu}] ,
\label{action_Einstein}
 \end{equation}
where $G$ is the gravitational constant, 
${\cal R}$ is the Ricci scalar with respect to the metric ${g_{\mu\nu}}$, 
and $\Phi$ is the scalar field. 
The term $S_m$ denotes the action of additional matter fields 
which are represented by $\Psi_{m}$. 

This action in the Einstein frame is obtained from the Jordan frame action 
by a conformal transformation of the metric 
${\tilde g}_{\mu\nu} = {A}^{2}(\varphi)g_{\mu\nu}$
with a coupling function $A(\varphi)$.
Here we have restricted to the case with no scalar potential. 
In addition, in order to satisfy the weak equivalence principle, 
we demand that
in the physical Jordan frame
the scalar field does not couple directly 
to the additional matter fields. 
Note, that in the following we set $c=G=1$.

By a variation of the action (\ref{action_Einstein}) with respect 
to the fields, we get the coupled set of field equations in the Einstein frame.
The Einstein equations read
\begin{equation} 
{\cal R}_{\mu\nu} - \frac{1}{2}g_{\mu\nu}{\cal R} =
  2\partial_{\mu}\varphi \partial_{\nu}\varphi   -
g_{\mu\nu}g^{\alpha\beta}\partial_{\alpha}\varphi
\partial_{\beta}\varphi
+ 8\pi T_{\mu\nu} ,
\end{equation}
where ${\cal R}_{\mu\nu}$ is the Ricci tensor, 
and $T_{\mu\nu}$ is the stress-energy tensor 
of the matter content of the action (\ref{action_Einstein}). 
The scalar field equation is given by
\begin{equation}
 \nabla^{\mu}\nabla_{\mu}\varphi = - 4\pi k(\varphi)T,
\end{equation}  
where $T = T^{\mu}_{\mu}$, 
and $k(\varphi)= \frac{d\ln({A}(\varphi))} {d\varphi}$ 
is the logarithmic derivative of the coupling function $A(\varphi)$, 
\begin{equation}
{A}(\varphi)=e^{\frac{1}{2}\beta\varphi^2} \ , \ \ \ k(\varphi)=\beta \varphi .
\label{A}
\end{equation}
For a massless scalar field
the coupling constant $\beta$ is strongly constrained 
by observations of the binary pulsar PSR J1738+0333 \cite{Freire:2012mg}, 
requiring
\begin{equation}
\frac{d^{2} \ln({\cal  A}(\varphi))}{d\varphi^{2}}|_{\varphi=0}=\beta \ge -4.5 .
\end{equation}

We model the neutron stars in the physical Jordan frame
as a self-gravitating perfect fluid
with stress-energy tensor ${\tilde T}_{\mu\nu}$
\begin{equation}
 {\tilde T}_{\mu\nu}= (\tilde \varepsilon + \tilde p){\tilde u}_{\mu}
{\tilde u}_{\nu} + {\tilde p} {\tilde g}_{\mu\nu} , 
\end{equation}
where $\tilde \varepsilon $, $\tilde p$ and $\tilde u$ 
represent the energy density,
the pressure and the four-velocity in the Jordan frame, respectively,
i.e., $T^\mu_{\ \nu} = A^4(\varphi) \tilde T^\mu_{\ \nu}$.

The following form of the metric is then used
to construct the static spherically symmetric neutron star models
in the Einstein frame
\begin{equation}
ds^{2}=-e^{2\nu(r)}dt^{2}+e^{2\lambda(r)}d r^{2} + r^{2} (d \theta^{2} 
+\sin^{2}\theta d \phi^{2}),
\label{eq:metric}
\end{equation}
where the metric functions $\nu$ and $\lambda$ depend only 
on the radial coordinate $r$.  
The coupled set of field equations for the neutron star models then
reduces to a system of ordinary differential equations (ODEs). 
In order to integrate this set of ODEs, 
an EOS in the form 
$\tilde \varepsilon=\tilde \varepsilon(\tilde p)$
is needed, which will be addressed in subsection \ref{sec:EOS}.

\subsection{QNMs: general formalism and axial perturbations}

The study of GW mode spectra of compact stars is expected to
provide crucial information about the internal structure of the stars. 
Here we focus on the \textit{axial} modes. 
These partity-odd modes of the metric do not couple to the fluid,
and thus the fluid does not pulsate.
In particular, we perturb the static background models 
constructed in our previous paper \cite{Motahar:2017blm}.

We consider linear non-radial perturbations of the metric Ansatz 
and the fluid \cite{Thorne:1967a},
but allow also for perturbations of the scalar field.
The perturbations of the metric and the four-velocity read
\begin{eqnarray}
g_{\mu\nu} = g_{\mu\nu}^{(0)} + \epsilon h_{\mu\nu} , \\
\tilde{u}_{\mu} = \tilde{u}_{\mu}^{(0)} + \epsilon \delta \tilde{u}_{\mu} ,
\end{eqnarray}
where $\epsilon<<1$ is the perturbation parameter. 
At zeroth order we have the static spherically symmetric solution 
given by the metric $g_{\mu\nu}^{(0)}$ and
the fluid four-velocity $\tilde{u}_{\mu}^{(0)}$. 
At first order in the perturbation parameter $\epsilon$ 
we obtain the perturbation of the metric $h_{\mu\nu}$ 
and the four-velocity $\delta \tilde{u}_{\mu}$.

We expand in tensorial spherical harmonics \cite{Thorne:1980ru}, 
introducing the multipole numbers $l$ and $m$. 
Then the perturbations split into the two separate classes 
of axial and polar perturbations. 
Axial perturbations transform as $(-1)^{l+1}$ under parity transformations,
therefore they do not couple to scalar perturbations. 
Thus they do not perturb the pressure, energy density or scalar field.
The latter perturbations would appear only in the polar modes,
which are parity-even and transform as $(-1)^{l}$.

Axial perturbations are described by the metric perturbations
\begin{equation}
h_{\mu\nu}^{(axial)} =  \sum\limits_{l,m}
\left[
\begin{array}{c c c c}
	0 & 0 & - h_0	\frac{1}{\sin\theta}\frac{\partial}{\partial\phi}Y_{lm} 
	& h_0	\sin\theta\frac{\partial}{\partial\theta}Y_{lm} \\
  0 & 0 & - h_1	\frac{1}{\sin\theta}\frac{\partial}{\partial\phi}Y_{lm} 
  & h_1\sin\theta\frac{\partial}{\partial\theta}Y_{lm} \\
	-h_0\frac{1}{\sin\theta}\frac{\partial}{\partial\phi}Y_{lm} & 
	- h_1\frac{1}{\sin\theta}\frac{\partial}{\partial\phi}Y_{lm}  & 0 & 0 \\
	h_0	\sin\theta\frac{\partial}{\partial\theta}Y_{lm} & 
	h_1	\sin\theta\frac{\partial}{\partial\theta}Y_{lm} & 0 & 0
\end{array}
\right] ,
\label{pert1}
\end{equation}
and the four-velocity perturbations
\begin{equation}
\delta \tilde{u}_{\mu}^{(axial)}= 0.
\label{pert2}
\end{equation}
The perturbation functions $(h_0, h_1)$ depend in general 
on the radial coordinate $r$, time $t$ and the multipole
numbers $l$ and $m$. 

Introducing the axial perturbations (\ref{pert1})-(\ref{pert2})
into the field equations, 
the minimal set of equations describing the perturbation is
\begin{equation}
-\partial^{2}_{t}h_{1}+ (\partial_{r}-\frac{2}{r})\partial_{t}h_{0} - \dfrac{(l+2)(l-1)}{r^{2}}e^{2\nu} h_{1}= 0 ,
\end{equation}
\begin{equation}
- e^{2\lambda} \partial_{t}h_{0} + e^{2\nu} (\partial_{r}h_{1} + (\partial_{r}\nu - \partial_{r}\lambda)h_{1})= 0 ,
\end{equation}
Now for simplicity we introduce a new function $X$ via 
$h_1 = r e^{\nu-\lambda}X$. 
Then the relevant equation can be expressed in terms of 
a second order ODE for $X=\frac{h_1 e^{\lambda-\nu }}{r}$,
\begin{equation}
\frac{\partial^2 X}{\partial t^2} 
-e^{\nu-\lambda} \frac{\partial}{\partial r}\left[e^{\nu-\lambda}\dfrac{\partial X}{r}\right]
+ e^{2\nu}\left[ \dfrac{l(l+1)}{r^{2}}-\dfrac{3}{r^{2}}(1-e^{-2\lambda})
+ 4\pi {\cal A}^{4}(\tilde{\varepsilon}-\tilde{p})\right] X =0 ,
\end{equation}
representing the generalized Regge-Wheeler equation 
for STT neutron stars, where we have also used 
\begin{equation}
-r e^{\nu-\lambda}\dfrac{d}{dr}\left(\dfrac{e^{\nu-\lambda}}{r^{2}}\right)= \dfrac{2}{r^{2}}-\dfrac{3}{r^{2}}(1-e^{-\lambda})+4\pi {\cal A}^{4}(\tilde{\varepsilon}-\tilde{p}).
\end{equation} 
In the limit $\beta=0$, the classical Regge-Wheeler equation 
for neutron stars in GR is obtained. 
In the limit $\tilde{p}=\tilde{\varepsilon}=0$, 
the equation for axial perturbations of static black holes is recovered. 

Assuming a harmonic time-dependence, $X(r,t)=X(r)e^{-i\omega t}$,
the time-independent equation becomes a Schr\"odinger-type equation,
\begin{equation}
 -e^{\nu-\lambda} \frac{\partial}{\partial r}\left[e^{\nu-\lambda}\dfrac{\partial X}{r}\right]
+\left[\omega^{2}+ e^{2\nu}\left( \dfrac{l(l+1)}{r^{2}}-\dfrac{3}{r^{2}}(1-e^{-2\lambda})
+ 4\pi {\cal A}^{4}(\tilde{\varepsilon}-\tilde{p})\right)\right] X =0 .
\end{equation}
The eigenfrequency $\omega$ is a complex number: 
$\omega=\omega_{R}+i\omega_{I}$. 
The real part $\omega_{R}$ corresponds to the frequency of the modes,
and the imaginary part $\omega_{I}$ 
is the inverse of the damping time $\tau$, $\omega_{I}=1/\tau$. 

The perturbation should be regular at the center of the star, 
i.e., the solution should be non-divergent there. 
Therefore, the Regge-Wheeler function should behave as
\begin{equation}
X = X_{l+1}r^{l+1} + O(r^{l+3})
\end{equation}
in both theories considered, GR and STT.

In general a solution of the (generalized)
Regge-Wheeler equation is given 
by a superposition of an incoming signal $X^{in}$ 
and an outgoing signal $X^{out}$,
\begin{equation}
\lim_{r\to\infty}X^{in} \sim e^{i\omega r}, \lim_{r\to\infty}X^{out} \sim e^{-i\omega r}. 
\end{equation}
Since we would like to find the resonant frequencies 
and damping times of GWs coming out of the star, 
we need to choose as boundary condition
to have a purely outgoing wave at radial infinity.

In our numerical study we will focus on the $l=2$ curvature modes.
We note, that the frequency of the modes and the damping times 
of the QNMs are the same in the physical Jordan frame 
and in the Einstein frame. 
 
\subsection{Equation of State}\label{sec:EOS}

We here employ 13 realistic EOSs, and in addition
a polytropic EOS for comparison.
The polytropic EOS is given by
\begin{equation}
\tilde \varepsilon = 
K \frac{{\tilde \rho}^\Gamma}{\Gamma - 1} + {\tilde \rho} \ , \ \ \
\tilde p=  K {\tilde \rho}^\Gamma \ , \ \ \
\Gamma =  1 + \frac{1}{N} ,
\label{poly}
\end{equation}
where $\tilde \rho$ is the baryonic mass density, 
and we have chosen the polytropic constant $K =1186.0$, 
and the polytropic index $N=0.7463$ for the adiabatic index $\Gamma$.

The 13 realistic EOSs are obtained from effective models 
of the nuclear interactions subject to different assumptions.
In order to compare the effects of exotic matter 
in the properties of the configurations, 
we have studied two EOSs containing only nuclear matter: 
SLy \cite{Douchin:2001sv} and APR4 \cite{Akmal:1998cf}. 
For EOSs containing nucleons and hyperons 
we have considered five EOSs: 
BHZBM \cite{Bednarek:2011gd}, GNH3 \cite{Glendenning:1984jr}, 
H4 \cite{Lackey:2005tk} and WCS1, WSC2 \cite{Weissenborn:2011ut}. 
For hybrid matter consisting of quarks and nucleons 
we have employed three EOSs: 
ALF2, ALF4 \cite{Alford:2004pf}, 
and WSPHS3 \cite{Weissenborn:2011qu}.

We have also employed three new EOSs proposed  recently
by Paschalidis et al.\cite{Paschalidis:2017qmb},
who have developed new parameterizations of hybrid quark-hadron EOSs 
allowing for a third family of stable compact objects.
The hybrid stars (HSs) interiors have a single-phase quark core 
enclosed by a hadronic shell with a first order 
quark-hadron phase transition at their interface. 
In particular, we consider three EOSs from two different subsets 
of \cite{Paschalidis:2017qmb}.
From the subset labled ''ACS-I'' we choose the EOS
with j=0.43 and transition pressure 
$\tilde p_{tr}=1.7 \times 10^{35}$ dyn cm$^{-2}$ 
at $\tilde \varepsilon_{tr}=8.34 \times 10^{14}$  g   cm$^{-3}$, 
and $ C_{s}^{2} = 0.8$. 
From the subset labled ''ACS-II'' 
we choose two EOSs with $j=0.8$ and $j=1.0$ with transition pressure 
$\tilde p_{tr}=8.34 \times 10^{34}$ dyn cm$^{-2}$ 
at $\tilde \varepsilon_{tr}=6.58 \times 10^{14}$ g cm$^{-3}$, 
and $ C_{s}^{2} = 1$.
Interestingly, it has been claimed in \cite{Paschalidis:2017qmb},
that GW170817 can constrain the properties of hybrid stars (HSs),
and that GW170817 is consistent with the coalescence 
of a hybrid star-neutron star binary.

All the EOSs considered here are consistent with 
the observed neutron star mass of about $2 M_{\odot}$
of the candidate pulsars PSR J1614-2230 \cite{Demorest:2010bx} 
and PSR J0348+0432 \cite{Antoniadis:2013pzd}.  
We note, that recent constraints on neutron star masses and radii 
have been discussed in \cite{Lattimer:2013hma,Ozel:2016oaf,Most:2018hfd}.
Even more, the first steps into constraining the EOS combining GW detection with electromagnetic observations have already been done \cite{Radice:2017lry, Tews:2018mr, Annala:2018gkv, Coughlin:2018etal, TheLIGOScientific:2018eos1}, showing that with more detections of NS mergers the matter composition of the star could be greatly constrained.

\subsection{Numerical Method}\label{sec:Numerical Method}

By solving the stellar structure equations 
with appropriate boundary conditions satisfying regularity at the center 
and asymptotic flatness, we have obtained numerically 
the configurations of static neutron stars.
For the numerical integration
we have used the ODE solver package COLSYS \cite{Ascher:1979iha}. 

To implement the EOSs in the numerical code,
we have used different methods.
The analytic relation for the relativistic polytrope 
has been simplest to implement.
For the equations ALF2, ALF4, APR4, GNH3, H4 and SLy
we have implemented a piecewise polytropic interpolation,
where different regions of the EOS are approximated 
by specific polytropes \cite{Read:2008iy}.
For the tabulated EOSs BHZBM, WCS1, WCS2 and WSPHS3 
we have used a piecewise monotonic cubic Hermite interpolation 
of the data points. 
However, for the EOSs ACS-I j=0.43, ACS-II j=0.8 and ACS-II j=1.0,
which have jumps due to the phase transitions,
we have used a linear interpolation of the data points.

\section{Results} \label{sec:Results}

We now present our results for axial QNMs of static neutron stars 
for the 13 realistic EOSs and the polytropic EOS in STT, 
employing the coupling function
$A=e^{\frac{1}{2}\beta\varphi^2}$ with $\beta=-4.5$.
We note that the GR solutions are included for vanishing scalar field. 
After recalling the neutron star models, we first show the
frequencies and damping times. Then we discuss 
several universal relations for the modes.

\subsection{Neutron star models}  \label{sec:Models}

\begin{figure}[t!]
\begin{center}
\mbox{\hspace{0.2cm}
\includegraphics[height=.45\textheight, angle =-90]{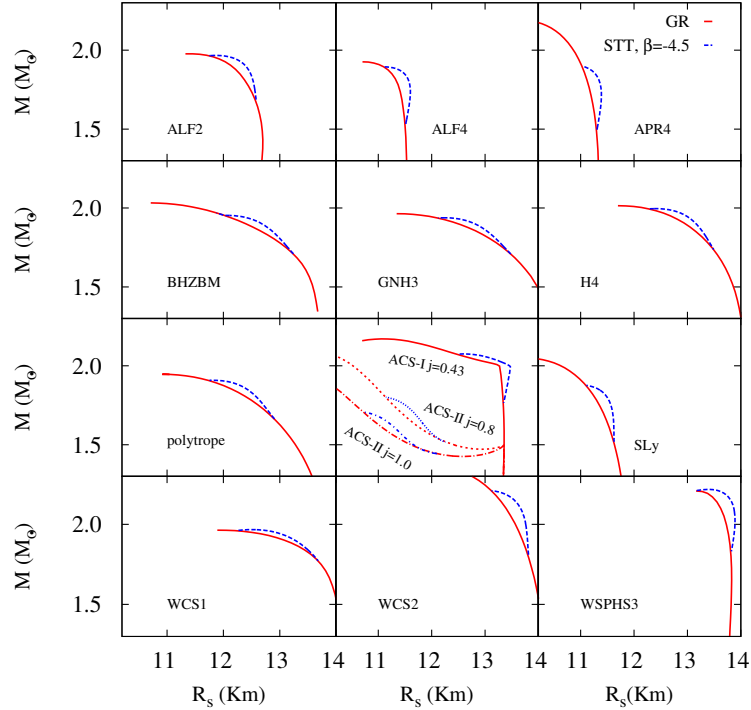}
}
\end{center}
\vspace{-0.5cm}
\caption{
Total mass $M$ in solar masses $M_\odot$
versus the physical radius $R_s$ in km 
of the neutron star models for all the EOSs considered
for GR configurations (red) and the scalarized solutions
for $A=e^{\frac{1}{2}\beta\varphi^2}$
with $\beta=-4.5$ (blue).
}
\label{plot_M_R_multi}
\end{figure}

\begin{figure}[p!]
\begin{center}
\vspace*{-0.5cm}
{\includegraphics[height=.45\textheight, angle =-90]{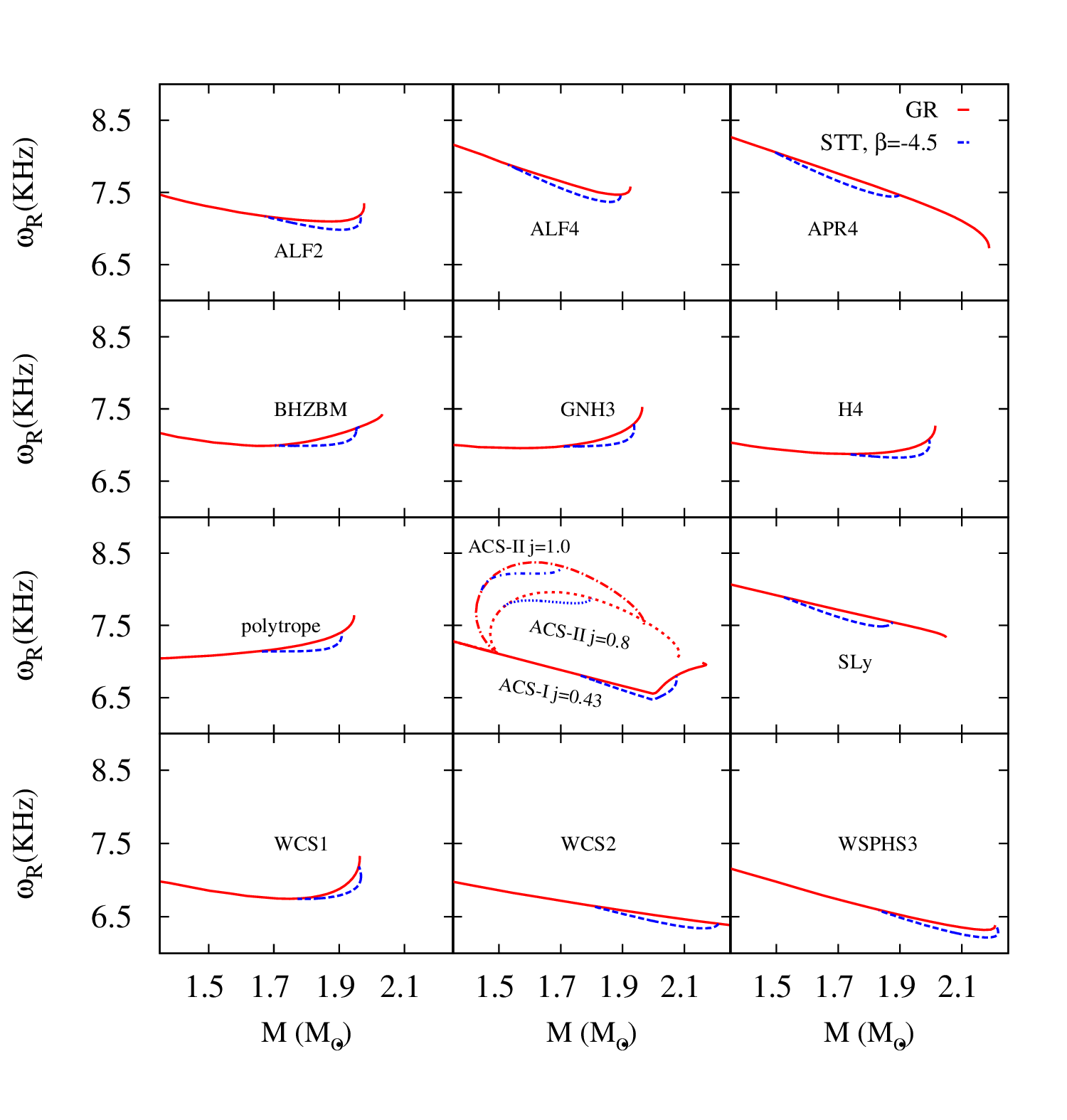}

\vspace*{-0.7cm}\hspace*{-10.5cm} (a)}

{\includegraphics[height=.45\textheight, angle =-90]{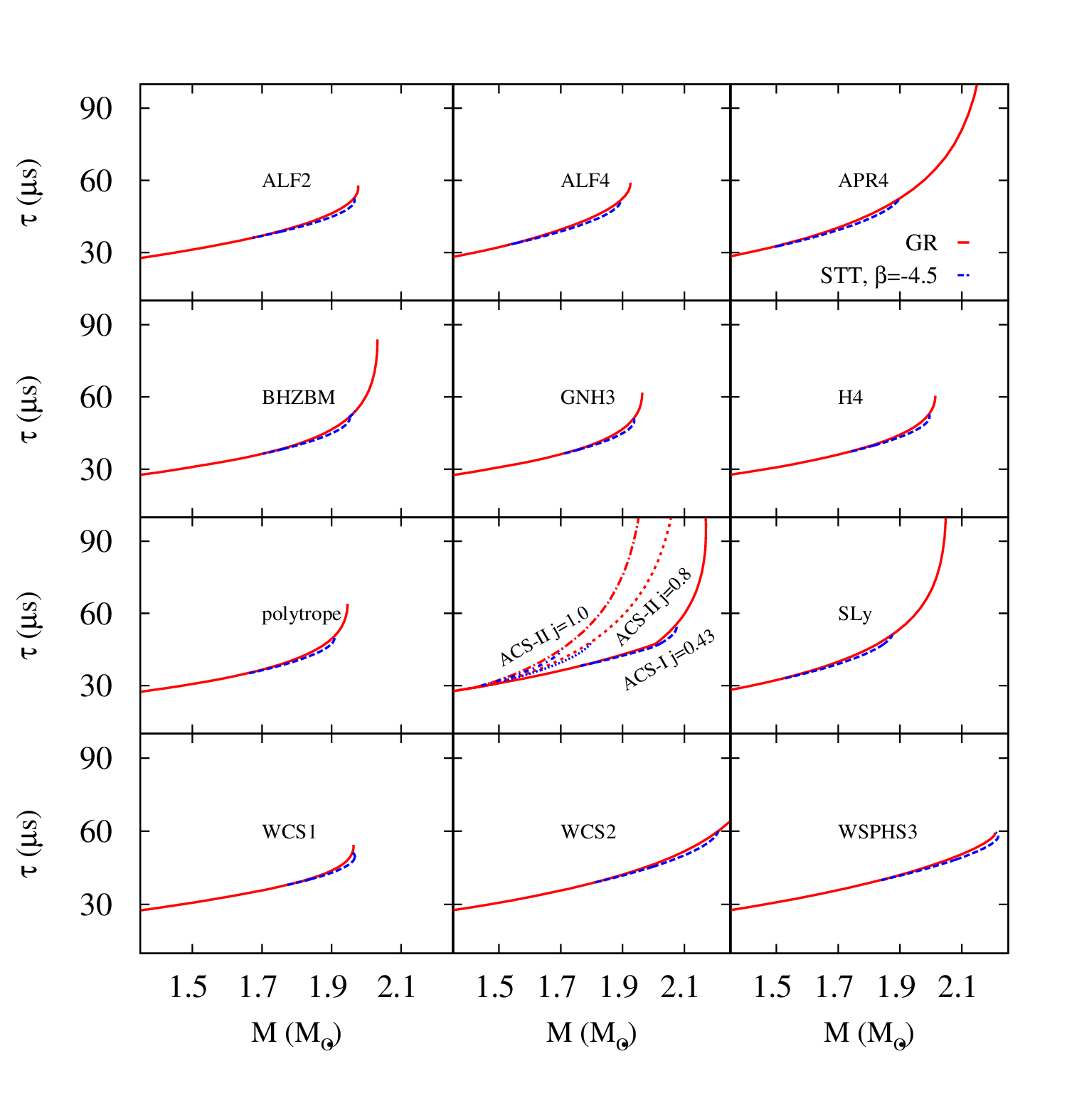}

\vspace*{-0.7cm}\hspace*{-10.5cm} (b)}
\end{center}
\caption{
Frequency $\omega_{R}$ in kHz (a) 
and damping time $\tau$ in $\mu$s (b)
versus 
the total mass $M$ in solar masses $M_\odot$
of the neutron star models for all the EOSs considered
for GR configurations (red) and the scalarized solutions
for $A=e^{\frac{1}{2}\beta\varphi^2}$ 
with $\beta=-4.5$ (blue).
}
\label{plot_omegaR_M_multi}
\end{figure}

To set the stage we present the neutron star models for all EOSs by
showing the total mass $M$ in solar masses $M_\odot$
versus the physical radius $R_s$ in km 
in Fig.~\ref{plot_M_R_multi},
where the scalarized models (blue) are compared to the GR models (red).
The mass--radius curves show that scalarized neutron star models
have typically slightly larger mass in the radius range 
where the spontaneous scalarization arises.

Note that the mass--radius curves for the 
hybrid EOSs ACS-I j=0.43, ACS-II j=0.8, and ACS-II j=1.0 
exhibit sharp bends at specific points, 
which arise due to the jumps in the energy density when the phase transitions occur.
This feature appears also in the scalarized model 
with EOSs ACS-I. In the EOSs ACS-II models scalarization
arises only after the phase transition.

In our previous analysis of STT neutron star models
\cite{Motahar:2017blm} we confirmed that compactness ${\cal C}=M/R_s$
features prominently in various universal relations 
\cite{Yagi:2016bkt,Doneva:2017jop}, 
including, in particular, the $I$--${\cal C}$ relations
between a scaled moment of inertia $I$ and compactness ${\cal C}$
\cite{Staykov:2016mbt,Lattimer:2004nj,Breu:2016ufb}.
As a new universal feature we found a relation between 
the scalarization and the gravitational potential at the center of the star.
We note that also the EOSs ACS-I and ACS-II employed here do satisfy these
universality relations, i.e., the phase transitions of these EOSs
do not spoil these universal relations.

\subsection{QNMs and universal relations} \label{sec:Universal}

Turning now the discussion to the axial QNMs, we exhibit
in Fig.~\ref{plot_omegaR_M_multi}(a)
and Fig.~\ref{plot_omegaR_M_multi}(b)
the frequency $\omega_R$ in kHz and the damping time $\tau=1/\omega_I$
im $\mu$s, respectively,
versus the total mass $M$ in solar masses $M_\odot$
for the fundamental $l=2$ mode.
For all EOSs considered, the frequencies are found in the range of 6 to 9 kHz. 
The effect of scalarization is to reduce the frequency 
as compared to the non-scalarized GR configurations
with a similar mass.
Also the damping times are slightly decreased in scalarized
configurations.

Let us next address universal relations for the neutron star QNMs.
We note that universal relations between the
scaled frequency and compactness as well as the scaled damping time
and compactness have been proposed
\cite{Andersson:1996pn,Andersson:1997rn,Kokkotas:1999mn,Benhar:1998au,Benhar:2004xg,Tsui:2004qd,BlazquezSalcedo:2012pd,Blazquez-Salcedo:2013jka}.
These universal relations for GR neutron star models include relations for
the $f$-mode, the polar $w$-modes and the axial $w$-modes.
These relations can then be exploited to learn about the
properties of the neutron stars and their EOS, once
the modes have been measured.

In the following we consider three universal relations for the
fundamental $l=2$ axial mode, starting with the dependence of the
frequency and damping time on the compactness.
Whereas first parametrizations scaled the frequency 
with the radius $R_s$ and the damping time with the mass $M$
\cite{Andersson:1997rn,Benhar:1998au},
later ones also considered a scaling of the frequency with the mass
\cite{Tsui:2004qd}. In particular,
Tsui and Leung \cite{Tsui:2004qd} proposed that such a scaling would
be appropriate for the $f$-mode, the polar $w$-modes and the axial $w$-modes,
demonstrating the validity also for the 1st excited polar and axial $w$-modes.

In their analysis Tsui and Leung \cite{Tsui:2004qd} concluded,
that the reason for the universality is lying in the mathematical structure 
of the axial and polar QNM equations, and in the fact that the
mass distribution function $m(r)/M$ is a simple polynomial in the radius,
characterized only by the compactness.
They expressed concern, however, for the case of a discontinuous 
or rapidly varying EOS.

\begin{figure}[p!]
\begin{center}
\vspace*{-0.5cm}
{\includegraphics[height=.45\textheight, angle =-90]{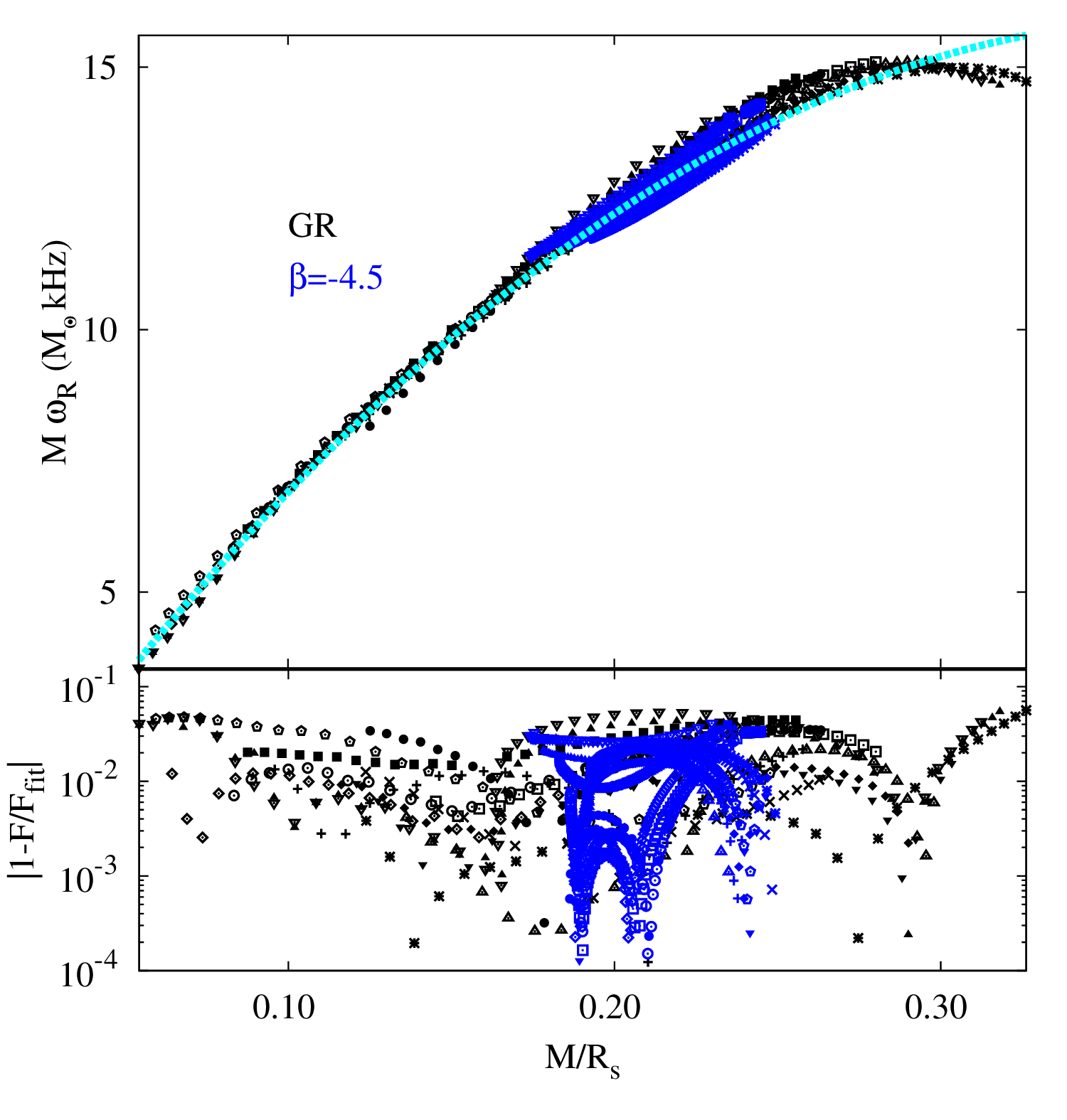}

\vspace*{-0.7cm}\hspace*{-10.5cm} (a)}

{\includegraphics[height=.45\textheight, angle =-90]{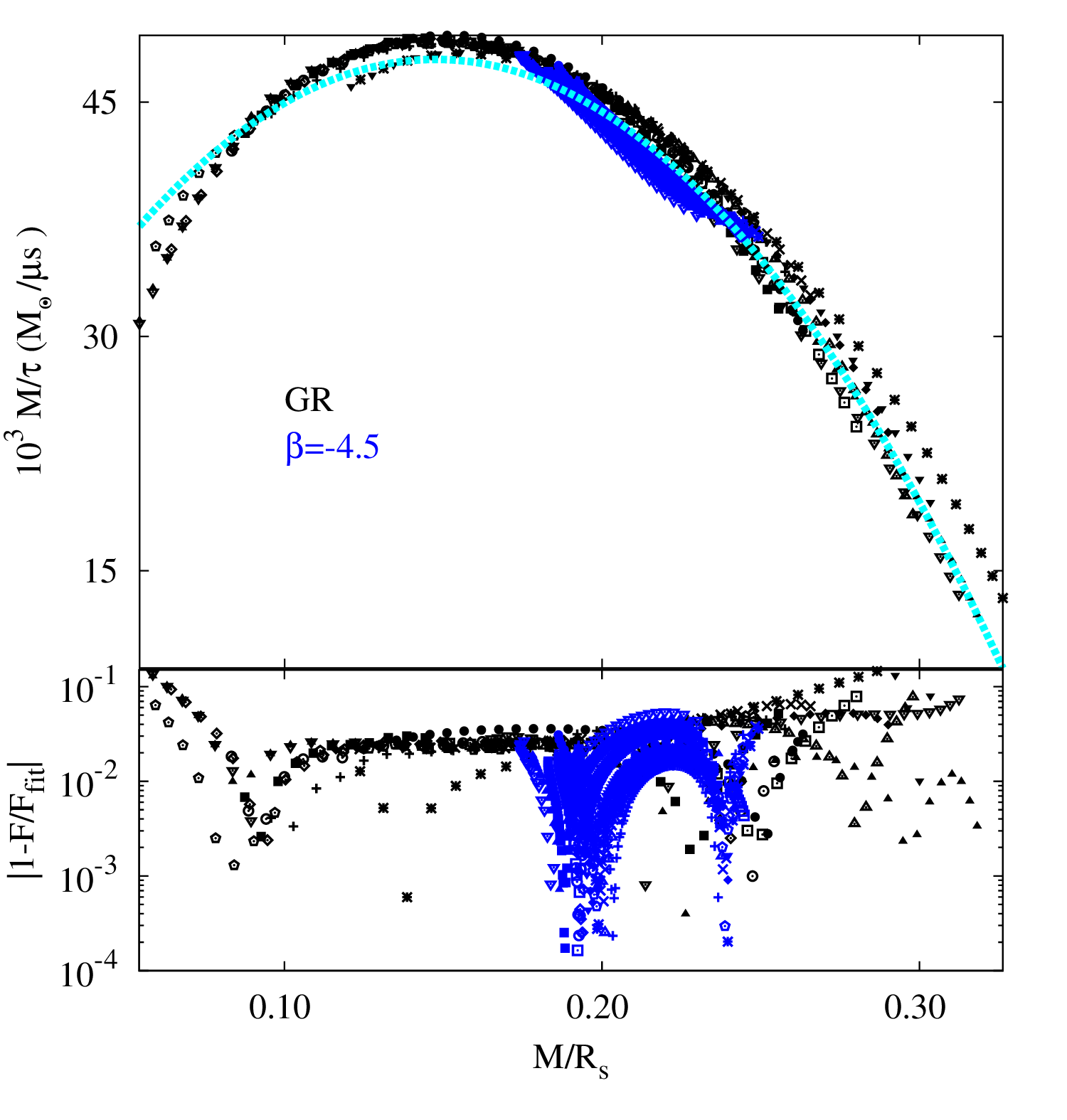}

\vspace*{-0.7cm}\hspace*{-10.5cm} (b)}
\end{center}
\vspace{-0.5cm}
\caption{
Frequency $\omega_{R}$ in kHz (a)
and damping time $\tau$ in $\mu$s (b) 
both scaled by the mass in $M_\odot$
versus
the compactness ${\cal C}=M/R_s$
of the neutron star models for all the EOSs considered
for GR configurations (black) and the scalarized solutions
for $A=e^{\frac{1}{2}\beta\varphi^2}$
with $\beta=-4.5$ (blue).
The  upper panels show the scaled values (symbols)
together with the fitted curves (dotted) of the universal relations.
The lower panels show the deviations
from the fits, $|1 - F/F_{\rm fit}|$. 
}
\label{plot_MR_omegaR_scaled_v3_w_error}
\end{figure}

In Fig.~\ref{plot_MR_omegaR_scaled_v3_w_error}(a) and (b)
we present the universal relation for the scaled frequency 
$M \omega_{R}$ --compactness ${\cal C}=M/R_s$
and the universal relation for the scaled damping time
$10^{3}M / \tau$ --compactness ${\cal C}$, respectively.
The blue symbols indicate the scalarized neutron star models, 
while the black symbols show the corresponding GR models.
Note, that for the GR case we have included only configurations 
up to the maximum mass.
The dotted cyan curves show the best fits 
according to Eqs.~(\ref{empiricalfrec})-(\ref{empiricalomgtau}) 
and table~\ref{tab:universal fits},
including the scalarized models.
The lower panels in the figures exhibit the deviations
(in blue for STT and in black for GR)
from the fitted values, $|1 - F/F_{\rm fit}|$,
which are always below $10\%$.

The fits of the frequency and the damping time are performed
with polynomials quadratic in the compactness,
\begin{eqnarray}
\omega(kHz) = 
\frac{1}{M(M_\odot)}\left[a_1\left(\frac{M}{R_s}\right)^{2} + b_1\frac{M}{R_s} + c_1\right], 
\label{empiricalfrec}
\\
\frac{10^3}{\tau(\mu s)} = 
\frac{1}{M(M_{\odot})}\left[a_2\left(\frac{M}{R}\right)^{2} + b_2\frac{M}{R} + c_2\right].
\label{empiricalomgtau}
\end{eqnarray}    
The constants $a_i$, $b_i$ and $c_i$, $i=1$,2, 
are given in table~\ref{tab:universal fits}.
We note that although the three EOS ACS-I $j=0.43$, ACS-II $j=0.8$, 
and ACS-II $j=1.0$ have jumps due to their phase transitions, 
they are perfectly fitting the universal relations.
Thus the phase transitions do not destroy the universality.

\begin{table}
\centering
\begin{tabular}{|c|c|c|c|c||}
\hline
$i$ & $a_i$ & $b_i$ & $c_i$ \\ 
\hline\hline 
\textbf{1} & $-115.6        \pm 1.8       (1.6\%)$ & $87.74                \pm 0.70 (0.80\%) $ & $-0.716 \pm 0.069 (9.7\%)$ \\ 
\hline 
\textbf{2} & $-1222.2        \pm 8.3        (0.68\%)$ & $361.0          \pm 3.2 (0.89\%)$ & $ 21.07          \pm 0.32       (1.5\%)$ \\ 
\hline 
\textbf{3} & $0.02157      \pm 0.00015    (0.68\%)$ & $0.4265         \pm 0.0062 (1.5\%)$ & $-1.335         \pm 0.050      (3.8\%)$ \\ 
\hline 
\textbf{4} & $-52.21       \pm  0.74       (1.4\%)$ & $66.26          \pm 0.51       (0.77\%)$ & $-3.762         \pm 0.088      (2.3\%)$ \\ 
\hline 
\textbf{5} & $-987.5       \pm 4.4       (0.44\%)$ & $ 535.48        \pm 3.0       (0.56\%)$ & $-22.74          \pm 0.52      (2.3\%)$ \\ 
\hline 
\end{tabular} 
\caption{Fit parameters for the universal relations 
including all EOSs.}
\label{tab:universal fits}
\end{table}

\begin{figure}[h!]
\begin{center}
\mbox{\hspace{0.2cm}
\includegraphics[height=.45\textheight, angle =-90]{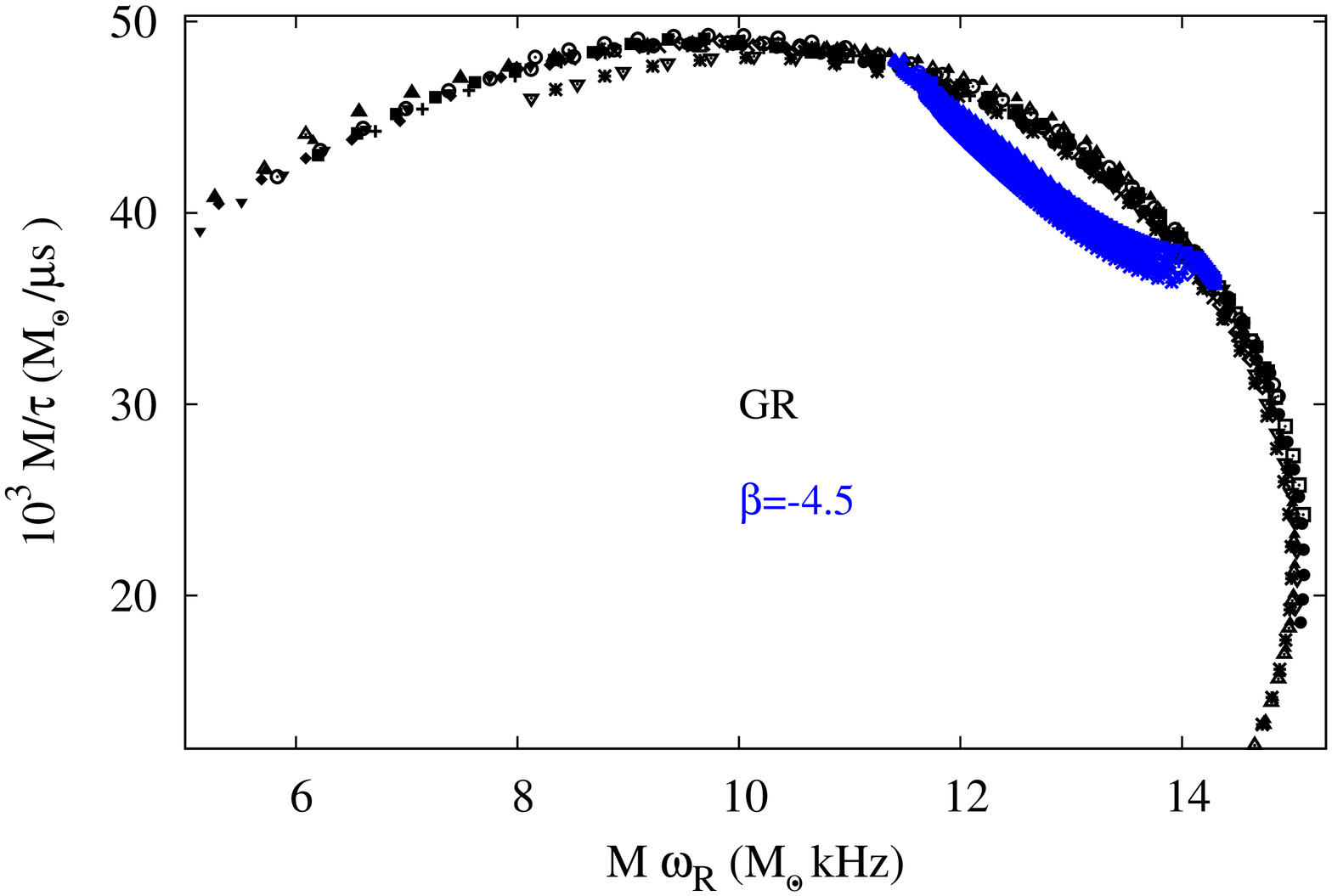}
}
\end{center}
\vspace{-0.5cm}
\caption{
The scaled quantity $10^{3}M / \tau$ (in $10^3 M_\odot$/$\mu$s) as a function of $M \omega_R$ (in $M_\odot$ kHz).
GR configurations are in black and the scalarized solutions
for $A=e^{\frac{1}{2}\beta\varphi^2}$
with $\beta=-4.5$ in blue.
}
\label{plot_MomegaR_MomegaI_noerror2}
\end{figure}

In Figure ~\ref{plot_MomegaR_MomegaI_noerror2} we show the scaled quantity $10^{3}M / \tau$ as a function of $M \omega_R$. Although the information in this Figure is the same as in the last two previous plots we have discussed, it is interesting to note that in this case the configurations in the scalar-tensor theory present a qualitative different behaviour than the GR configurations.

\begin{figure}[h!]
\begin{center}
\mbox{\hspace{0.2cm}
\includegraphics[height=.45\textheight, angle =-90]{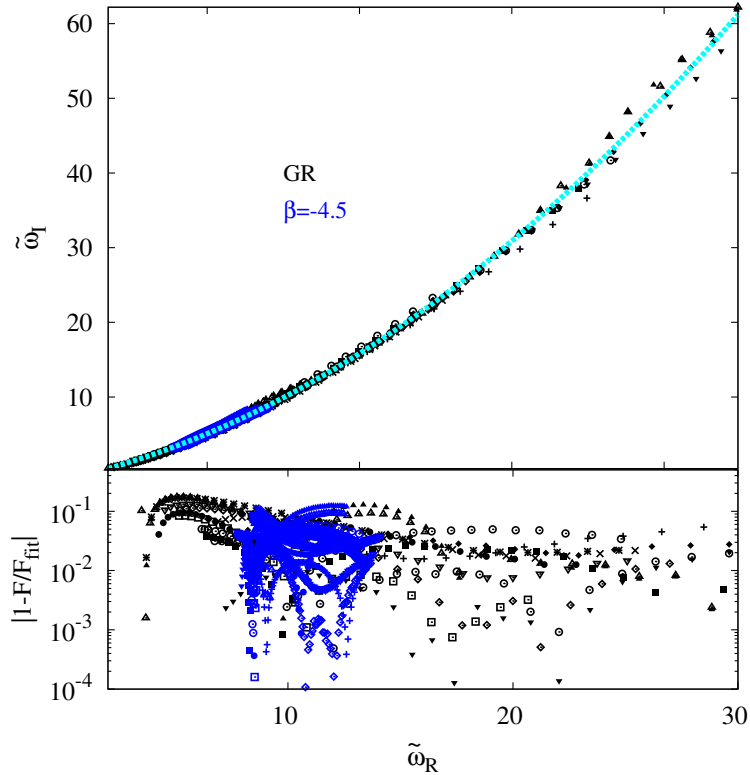}
}
\end{center}
\vspace{-0.5cm}
\caption{
Relation between the imaginary part of the eigenvalue $\omega$ with the real part, when both are normalized with respect to the central pressure of the neutron star ($\tilde \omega_I$ vs $\tilde \omega_R$). GR configurations are in black and the scalarized solutions
for $A=e^{\frac{1}{2}\beta\varphi^2}$
with $\beta=-4.5$ in blue.
The upper panel shows the scaled values (symbols)
together with the fitted curve (dotted) of the universal relation.
The lower panel shows the deviations
from the fit, $|1 - F/F_{\rm fit}|$.
}
\label{plot_omegaR_omegaI_v3_w_error3}
\end{figure}

It has been shown before \cite{BlazquezSalcedo:2012pd,Blazquez-Salcedo:2013jka,Blazquez-Salcedo:2015ets,Blazquez-Salcedo:2018tyn,Blazquez-Salcedo:2018qyy} that scaling the $\omega$ with respect to the central pressure also provides an almost universal relation. Defining the dimensionless quantities $\tilde{\omega}_R = \omega_R/\sqrt{\tilde p_c}$ and $\tilde{\omega}_I = \omega_I/\sqrt{\tilde p_c}$, we show in Fig.~\ref{plot_omegaR_omegaI_v3_w_error3} the relation between $\hat{\omega}_I$ and  $\hat{\omega}_R$.
This relation is almost the same for both GR and scalarized 
neutron star models, meaning the effect of changing the theory is less or equal to the effect of changing the equation of state. 
The dotted cyan curve shows the best fit 
according to Eq.~(\ref{pc_scaled_modes}) 
and table~\ref{tab:universal fits},
including the scalarized models.
The lower panel shows the deviations
from the fit.
Here a quadratic relation holds between the imaginary frequency part 
$\tilde \omega_I$ and the real part $\tilde \omega_R$,
\begin{eqnarray}
\tilde \omega_{I}=
a_3 \left(\tilde \omega_{R}\right)^{2}
+  b_3 \tilde \omega_{R}
+ c_3  
. \label{pc_scaled_modes}
\end{eqnarray}
The constants $a_i$, $b_i$ and $c_i$, $i=3$, 
are given in table~\ref{tab:universal fits}.

Lau et al.~\cite{Lau:2009bu} observed a universal relation
for the $f$-mode, where the scaled frequency 
is considered as a quadratic polynomial of the scaled moment of inertia.
We here present such a universal relation 
for the axial modes.
In particular, we scale the moment of inertia by the mass
and consider the quantity $\eta=\sqrt{M^{3}/I}$ 
(following \cite{Motahar:2017blm}). 

\begin{figure}[p!]
\begin{center}
\vspace*{-0.5cm}
{\includegraphics[height=.45\textheight, angle =-90]{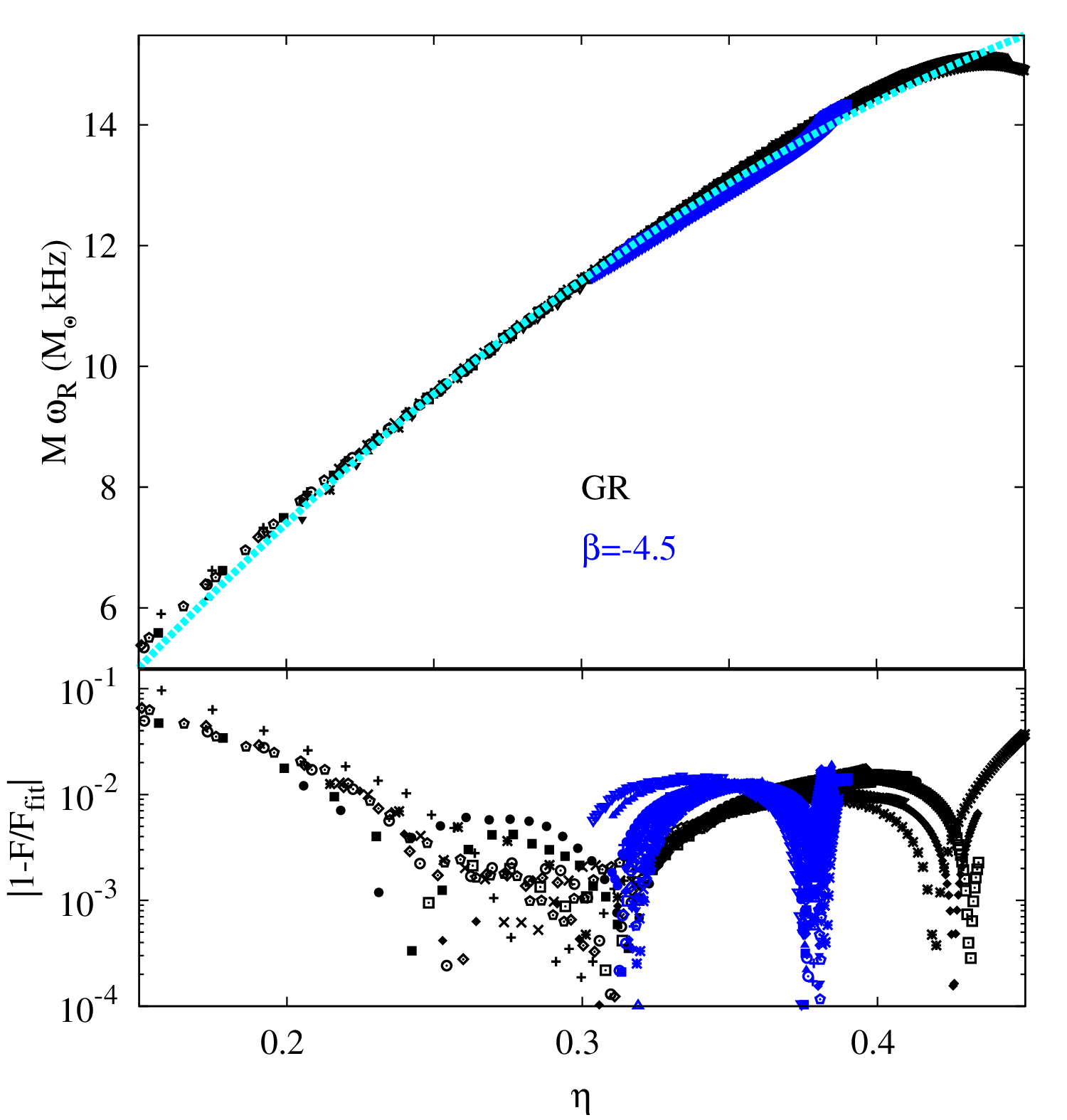}

\vspace*{-0.7cm}\hspace*{-10.5cm} (a)}

{\includegraphics[height=.45\textheight, angle =-90]{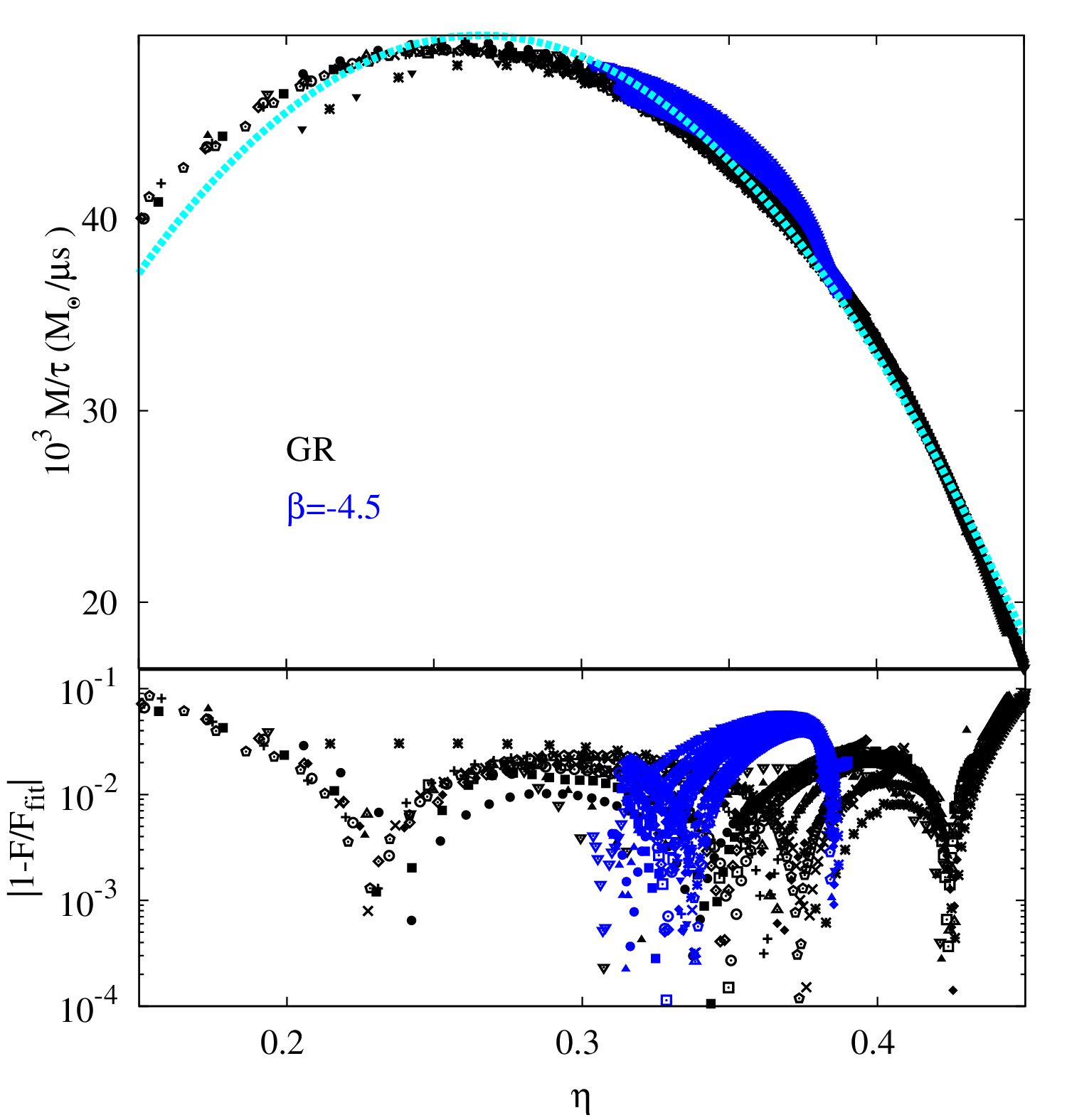}

\vspace*{-0.7cm}\hspace*{-10.5cm} (b)}
\end{center}
\vspace{-0.5cm}
\caption{
Frequency $\omega_{R}$ in kHz (a)
and inverse damping time $\tau$ in $\mu$s (b)
both scaled by the mass in $M_\odot$
versus
the scaled moment of inertia $\eta=\sqrt{M^{3}/I}$
of the neutron star models for all the EOSs considered
for GR configurations (black) and the scalarized solutions
for $A=e^{\frac{1}{2}\beta\varphi^2}$
with $\beta=-4.5$ (blue).
The  upper panels show the scaled values (symbols)
together with the fitted curves (dotted) of the universal relations.
The lower panels show the deviations
from the fits, $|1 - F/F_{\rm fit}|$.
}
\label{plot_MomegaR_eta_scaled_error4}
\end{figure}

The results for this universal relation
are shown in Fig.~\ref{plot_MomegaR_eta_scaled_error4} (a) and (b).
The dotted cyan curve in Fig.~\ref{plot_MomegaR_eta_scaled_error4} (a) shows the best fit 
according to Eqs.~(\ref{empiricalfrec2})-(\ref{empiricalomgtau2}) 
and table~\ref{tab:universal fits},
including the scalarized models. 
However, we considered only the GR solutions for the best fit in Fig.~\ref{plot_MomegaR_eta_scaled_error4} (b).
The lower panel shows the deviations
from the fits.%
The quadratic relations between 
 the scaled frequency $M \omega_R$ and scaled damping time $M/\tau$,
 with respect to the scaled moment of inertia
are given by
\begin{eqnarray}
\omega_R({\rm kHz}) = 
\frac{1}{M(M_\odot)}\left[a_4 \eta^{2} + b_4 \eta + c_4\right], 
\label{empiricalfrec2}
\\
\frac{10^3}{\tau(\mu s)} = 
\frac{1}{M(M_\odot)}\left[a_5 \eta^{2} + b_5 \eta + c_5\right].
\label{empiricalomgtau2}
\end{eqnarray}    
The constants $a_i$, $b_i$ and $c_i$, $i=4$, 5, 
are given in table~\ref{tab:universal fits}.

Thus we have confirmed and extended several universal relations 
between (scaled) quantities that are known in GR 
to this wide range of realistic EOSs in the scalarized neutron star models,
including EOSs with phase transitions.
In fact, the universal relations obtained in STT do not deviate 
from the GR relations significantly.

Depending on the observational data available, 
these relations could, for instance,
be used to approximately infer the compactness 
of a star, and once the compactness is known, 
the approximate mass of the star could be inferred. 

\section{Conclusions} \label{sec:Conclusions}

In this study, we have computed axial QNMs 
of static neutron stars in STT. For the matter
we have employed besides a polytropic EOS
a large number of realistic EOSs,
including nuclear, hyperonic and hybrid matter,
with three EOSs featuring phase transitions.

We have investigated the fundamental $l=2$ curvature mode 
for the scalarized neutron star models for all these EOSs
and compared the results to those of GR.
We have found that, in general, 
the frequency and the damping time of the modes
are higher in GR than in STT.

Our main concern, however, has been the study of universal relations
for these axial QNMs. 
Here we have confirmed and extended several universal relations
known in GR to this wide range of realistic EOSs,
considering both non-scalarized and scalarized neutron star models.
Interestingly, the universal relations including the scalarized
models do not deviate significantly from the respective GR relations.
All universal relations are satisfied at a better than 10\% level.
This holds also for the neutron star models based on an EOS
with a phase transition.

In particular, we have studied the universal relations
where i) the scaled frequency $M \omega_R$ and the scaled
inverse damping time $M/\tau$ are fitted by quadratic
polynomials in the compactness ${\cal C}=M/R_s$,
Eqs.~(\ref{empiricalfrec})-(\ref{empiricalomgtau}),
ii) the scaled $\omega_I/\sqrt{\tilde p_c}$ is fitted by a quadratic polynomial
in the scaled $\omega_R/\sqrt{\tilde p_c}$, 
Eq.~(\ref{pc_scaled_modes}),
and
iii) the scaled frequency $M \omega_R$ and the scaled
inverse damping time $M/\tau$ are fitted by quadratic
polynomials in $\eta= \sqrt{M^3/I}$,
involving the scaled moment of inertia $I$,
Eqs.~(\ref{empiricalfrec2})-(\ref{empiricalomgtau2}).

While these universal relations will not allow us to distinguish
between GR and scalarized STT neutron stars, since the differences
are far too small, the relations can be employed to extract
additional information on the properties of neutron stars.
For instance, if an axial mode were measured, one could extract
from the ratio $\omega_R/\omega_I$
between the frequency and the inverse damping time 
the compactness of the star \cite{Tsui:2004qd}. 
Subsequently, when the compactness is known, one could
use the universal relations 
(\ref{empiricalfrec})-(\ref{empiricalomgtau})
to read off the mass of the star and infer the radius via the
compactness.
Together they would yield valuable information on the EOS.

Similarly, from the ratio $\omega_R/\omega_I$
one could infer the quantity $\eta= \sqrt{M^3/I}$
involving the moment of inertia of the star
\cite{Lau:2009bu}.
Subsequently, when $\eta$ is known, one could
use the universal relations
(\ref{empiricalfrec2})-(\ref{empiricalomgtau2})
to read off the mass of the star 
and then determine the moment of inertia of the star.

As future work, we would like to study the polar modes 
of the neutron star models in STT, including the $f$-mode,
the $p$-modes, $g$-modes and $w$-modes.
It would also be interesting to study magnetized stars
and rapidly rotating stars.
While closer to reality they will, however, be more difficult to investigate.
 
\section*{Acknowledgment}
We would like to acknowledge support by the
DFG Research Training Group 1620 {\sl Models of Gravity}
and the COST Action CA16104 {\sl GWverse}.
JLBS would like to acknowledge support from the DFG project BL 1553. 
BK gratefully acknowledges support
from Fundamental Research in Natural Sciences
by the Ministry of Education and Science of Kazakhstan.


\begin{thebibliography}{99}

\bibitem{Abbott:2016blz} 
  B.~P.~Abbott {\it et al.} 
  Phys.\ Rev.\ Lett.\  {\bf 116}, 
  061102 (2016).

 \bibitem{Abbott:2016nmj} 
  B.~P.~Abbott {\it et al.} 
  Phys.\ Rev.\ Lett.\  {\bf 116}, 
  241103 (2016).

\bibitem{Abbott:2017vtc} 
  B.~P.~Abbott {\it et al.} 
  Phys.\ Rev.\ Lett.\  {\bf 118}, 
  221101 (2017).

\bibitem{Abbott:2017oio} 
  B.~P.~Abbott {\it et al.} 
  Phys.\ Rev.\ Lett.\  {\bf 119}, 
   141101 (2017).

\bibitem{TheLIGOScientific:2017qsa} 
  B.~P.~Abbott {\it et al.} 
  Phys.\ Rev.\ Lett.\  {\bf 119}, 
  161101 (2017).

\bibitem{Abbott:2017dke} 
  B. P. Abbott \textit{et} al., 
  Astrophys.\ J.\  {\bf 851}, 
  L16 (2017).

\bibitem{Abbott:2017gyy} 
  B.~P.~.Abbott {\it et al.} 
  Astrophys.\ J.\  {\bf 851}, 
  L35 (2017).

\bibitem{GBM:2017lvd} 
  B.~P.~Abbott {\it et al.} 
  Astrophys.\ J.\  {\bf 848}, 
  L12 (2017).

\bibitem{Coulter:2017wya} 
  D.~A.~Coulter {\it et al.},
  Science
  [Science {\bf 358}, 1556 (2017)]

\bibitem{Piro:2017zec} 
  A.~L.~Piro, B.~Giacomazzo and R.~Perna,
  Astrophys.\ J.\  {\bf 844}, no. 2, L19 (2017).

\bibitem{Thorne:1967a} 
  K.~S.~Thorne and A.~Campolattaro,
  Astrophys.\ J.\  {\bf 149}, 591 (1967).
  Erratum: [Astrophys.\ J.\  {\bf 152}, 673 (1968)].

\bibitem{Thorne:1969a}
  R.~Price and K.~S.~Thorne,
  Astrophys.\ J.\  {\bf 155}, 163 (1969).

\bibitem{Thorne:1969b}
  K.~S.~Thorne,
  Astrophys.\ J.\  {\bf 158}, 1 (1969).

\bibitem{Thorne:1969rba} 
  K.~S.~Thorne,
  Astrophys.\ J.\  {\bf 158}, 997 (1969).

\bibitem{Lindblom:1983ps} 
  L.~Lindblom and S.~L.~Detweiler,
  Astrophys.\ J.\ Suppl.\  {\bf 53}, 73 (1983).

\bibitem{Detweiler:1985zz} 
  S.~L.~Detweiler and L.~Lindblom,
  Astrophys.\ J.\  {\bf 292}, 12 (1985).

\bibitem{Chandrasekhar:1991fi} 
  S.~Chandrasekhar and V.~Ferrari,
  Proc.\ Roy.\ Soc.\ Lond.\ A {\bf 432}, 247 (1991).

\bibitem{Chandrasekhar:1991a} 
  S.~Chandrasekhar and V.~Ferrari,
  Proc.\ Roy.\ Soc.\ Lond.\ A {\bf 434}, 449 (1991).
  
\bibitem{Chandrasekhar:1991b} 
  S.~Chandrasekhar and V.~Ferrari,
  Proc.\ Roy.\ Soc.\ Lond.\ A {\bf 434}, 635 (1991).
  
\bibitem{Ipser:1991ind} 
  J.~R.~Ipser and R.~H.~Price,
  Phys.\ Rev.\ D {\bf 43}, 1768 (1991).
  
\bibitem{Kojima:1992ie} 
  Y.~Kojima,
  Phys.\ Rev.\ D {\bf 46}, 4289 (1992).

\bibitem{Kokkotas:2003mh} 
  K.~D.~Kokkotas and B.~F.~Schutz,
  Mon.\ Not.\ Roy.\ Astron.\ Soc.\  {\bf 255}, 119 (1992).

  \bibitem{Kokkotas:1994an} 
  K.~D.~Kokkotas,
  Mon.\ Not.\ Roy.\ Astron.\ Soc.\  {\bf 268}, 1015 (1994).

  \bibitem{Kokkotas:1999bd}
  K.~D.~Kokkotas and B.~G.~Schmidt,
  Living Rev.\ Rel.\  {\bf 2}, 2 (1999). 
  
 \bibitem{Nollert:1999ji} 
  H.~P.~Nollert,
  Class.\ Quant.\ Grav.\  {\bf 16}, R159 (1999).

 \bibitem{Ferrari:2007dd} 
  V.~Ferrari and L.~Gualtieri,
  Gen.\ Rel.\ Grav.\  {\bf 40}, 945 (2008).
 
\bibitem{Heiselberg:1999mq} 
  H.~Heiselberg and M.~Hjorth-Jensen,
  Phys.\ Rept.\  {\bf 328}, 237 (2000).
 
\bibitem{Haensel:2007yy} 
  P.~Haensel, A.~Y.~Potekhin and D.~G.~Yakovlev,
  Astrophys.\ Space Sci.\ Libr.\  {\bf 326} (2007).

\bibitem{Lattimer:2012nd} 
  J.~M.~Lattimer,
  Ann.\ Rev.\ Nucl.\ Part.\ Sci.\  {\bf 62}, 485 (2012).

\bibitem{Cowling:1941a}
  T.G.~Cowling,
  Mon.\ Not.\ Roy.\ Astron.\ Soc.\  {\bf 101}, 367 (1941).

\bibitem{Berti:2018cxi} 
  E.~Berti, K.~Yagi and N.~Yunes,
  Gen.\ Rel.\ Grav.\  {\bf 50}, 46 (2018)

\bibitem{Berti:2018vdi} 
  E.~Berti, K.~Yagi, H.~Yang and N.~Yunes,
  Gen.\ Rel.\ Grav.\  {\bf 50}, 49 (2018)

\bibitem{Capozziello:2011et}
  S.~Capozziello and M.~De Laurentis,
  Phys.\ Rept.\  {\bf 509}, 167 (2011)

\bibitem{Berti:2015itd}
  E.~Berti {\it et al.},
  Class.\ Quant.\ Grav.\  {\bf 32}, 243001 (2015)




\bibitem{Sotani:2009nm} 
  H.~Sotani,
  Phys.\ Rev.\ D {\bf 80}, 064035 (2009).
   
  \bibitem{Blazquez-Salcedo:2015ets} 
  J.~L.~Bl\'azquez-Salcedo, L.~M.~Gonz\'alez-Romero, J.~Kunz, S.~Mojica and F.~Navarro-L\'erida,
  Phys.\ Rev.\ D {\bf 93}, 
  024052 (2016).

\bibitem{Blazquez-Salcedo:2018tyn} 
  J.~L.~Bl\'azquez-Salcedo and K.~Eickhoff,
  Phys.\ Rev.\ D {\bf 97}, 104002 (2018)

\bibitem{Blazquez-Salcedo:2018qyy} 
  J.~L.~Bl\'azquez-Salcedo, D.~D.~Doneva, J.~Kunz, K.~V.~Staykov and S.~S.~Yazadjiev,
  arXiv:1804.04060 [gr-qc].
  

  

\bibitem{Brans:1961sx}
  C.~Brans and R.~H.~Dicke,
  Phys.\ Rev.\  {\bf 124}, 925 (1961).

\bibitem{Damour:1992we}
  T.~Damour and G.~Esposito-Farese,
  Class.\ Quant.\ Grav.\  {\bf 9}, 2093 (1992).

\bibitem{Fujii:2003pa} 
  Y.~Fujii and K.~Maeda,
  \textit{The Scalar-Tensor Theory of Gravitation},
  Cambridge University Press, Cambridge, (2003).

\bibitem{Damour:1993hw} 
  T.~Damour and G.~Esposito-Farese,
  Phys.\ Rev.\ Lett.\  {\bf 70}, 2220 (1993).

\bibitem{Damour:1996ke} 
  T.~Damour and G.~Esposito-Farese,
  Phys.\ Rev.\ D {\bf 54}, 1474 (1996).

\bibitem{Harada:1998ge} 
  T.~Harada,
  Phys.\ Rev.\ D {\bf 57}, 4802 (1998).

\bibitem{Salgado:1998sg}
  M.~Salgado, D.~Sudarsky and U.~Nucamendi,
  Phys.\ Rev.\ D {\bf 58}, 124003 (1998).

\bibitem{Sotani:2012eb} 
  H.~Sotani,
  Phys.\ Rev.\ D {\bf 86}, 124036 (2012).

\bibitem{Pani:2014jra} 
  P.~Pani and E.~Berti,
  Phys.\ Rev.\ D {\bf 90}, 024025 (2014).

\bibitem{Sotani:2017pfj} 
  H.~Sotani and K.~D.~Kokkotas,
  Phys.\ Rev.\ D {\bf 95}, 044032 (2017).

\bibitem{Doneva:2013qva} 
  D.~D.~Doneva, S.~S.~Yazadjiev, N.~Stergioulas and K.~D.~Kokkotas,
  Phys.\ Rev.\ D {\bf 88}, 084060 (2013).

\bibitem{Doneva:2014uma} 
  D.~D.~Doneva, S.~S.~Yazadjiev, N.~Stergioulas, K.~D.~Kokkotas and T.~M.~Athanasiadis,
  Phys.\ Rev.\ D {\bf 90}, 044004 (2014).

\bibitem{Doneva:2014faa} 
  D.~D.~Doneva, S.~S.~Yazadjiev, K.~V.~Staykov and K.~D.~Kokkotas,
  Phys.\ Rev.\ D {\bf 90}, 104021 (2014).
  
\bibitem{Staykov:2016mbt} 
  K.~V.~Staykov, D.~D.~Doneva and S.~S.~Yazadjiev,
  Phys.\ Rev.\ D {\bf 93}, 084010 (2016).

\bibitem{Yazadjiev:2016pcb} 
  S.~S.~Yazadjiev, D.~D.~Doneva and D.~Popchev,
  Phys.\ Rev.\ D {\bf 93}, 084038 (2016).

\bibitem{Doneva:2016xmf} 
  D.~D.~Doneva and S.~S.~Yazadjiev,
  JCAP {\bf 1611}, 019 (2016).

\bibitem{Sagunski:2017nzb} 
  L.~Sagunski, J.~Zhang, M.~C.~Johnson, L.~Lehner, M.~Sakellariadou, S.~L.~Liebling, C.~Palenzuela and D.~Neilsen,
  Phys.\ Rev.\ D {\bf 97}, no. 6, 064016 (2018)

  \bibitem{Sotani:2005qx}
  H.~Sotani and K.~D.~Kokkotas,
  Phys.\ Rev.\ D {\bf 71}, 124038 (2005).

\bibitem{Yagi:2016bkt}
  K.~Yagi and N.~Yunes,
  Phys.\ Rept.\  {\bf 681}, 1 (2017).

\bibitem{Doneva:2017jop}
  D.~D.~Doneva and G.~Pappas,
  arXiv:1709.08046 [gr-qc].

\bibitem{Andersson:1996pn} 
  N.~Andersson and K.~D.~Kokkotas,
  Phys.\ Rev.\ Lett.\  {\bf 77}, 4134 (1996)

\bibitem{Andersson:1997rn} 
  N.~Andersson and K.~D.~Kokkotas,
  Mon.\ Not.\ Roy.\ Astron.\ Soc.\  {\bf 299}, 1059 (1998)

\bibitem{Kokkotas:1999mn} 
  K.~D.~Kokkotas, T.~A.~Apostolatos and N.~Andersson,
  Mon.\ Not.\ Roy.\ Astron.\ Soc.\  {\bf 320}, 307 (2001).

\bibitem{Benhar:1998au} 
  O.~Benhar, E.~Berti and V.~Ferrari,
  Mon.\ Not.\ Roy.\ Astron.\ Soc.\  {\bf 310}, 797 (1999)
  [ICTP Lect.\ Notes Ser.\  {\bf 3}, 35 (2001)]
 
\bibitem{Benhar:2004xg} 
  O.~Benhar, V.~Ferrari and L.~Gualtieri,
  Phys.\ Rev.\ D {\bf 70}, 124015 (2004).

\bibitem{Tsui:2004qd}
  L.~K.~Tsui and P.~T.~Leung,
  Mon.\ Not.\ Roy.\ Astron.\ Soc.\  {\bf 357}, 1029 (2005)

\bibitem{BlazquezSalcedo:2012pd} 
  J.~L.~Bl\'azquez-Salcedo, L.~M.~Gonz\'alez-Romero and F.~Navarro-L\'erida,
  Phys.\ Rev.\ D {\bf 87}, 104042 (2013).
 
\bibitem{Blazquez-Salcedo:2013jka} 
  J.~L.~Bl\'azquez-Salcedo, L.~M.~Gonz\'alez-Romero and F.~Navarro-L\'erida,
  Phys.\ Rev.\ D {\bf 89}, 044006 (2014).
  
 \bibitem{Yagi:2013bca} 
  K.~Yagi and N.~Yunes,
  Science {\bf 341}, 365 (2013).

\bibitem{Will:2014kxa} 
  C.~M.~Will,
  Living Rev.\ Rel.\  {\bf 17}, 4 (2014)


\bibitem{Freire:2012mg} 
  P.~C.~C.~Freire {\it et al.},
  Mon.\ Not.\ Roy.\ Astron.\ Soc.\  {\bf 423}, 3328 (2012)

\bibitem{Motahar:2017blm} 
  Z.~Altaha Motahar, J.~L.~Bl\'azquez-Salcedo, B.~Kleihaus and J.~Kunz,
  Phys.\ Rev.\ D {\bf 96}, 
  064046 (2017).

\bibitem{Thorne:1980ru} 
  K.~S.~Thorne,
  Rev.\ Mod.\ Phys.\  {\bf 52}, 299 (1980).
 
\bibitem{Douchin:2001sv} 
  F.~Douchin and P.~Haensel,
  Astron.\ Astrophys.\  {\bf 380}, 151 (2001).

\bibitem{Akmal:1998cf} 
  A.~Akmal, V.~R.~Pandharipande and D.~G.~Ravenhall,
  Phys.\ Rev.\ C {\bf 58}, 1804 (1998).

\bibitem{Bednarek:2011gd} 
  I.~Bednarek, P.~Haensel, J.~L.~Zdunik, M.~Bejger and R.~Manka,
  Astron.\ Astrophys.\  {\bf 543}, A157 (2012).
  
\bibitem{Glendenning:1984jr} 
  N.~K.~Glendenning,
  Astrophys.\ J.\  {\bf 293}, 470 (1985).

\bibitem{Lackey:2005tk} 
  B.~D.~Lackey, M.~Nayyar and B.~J.~Owen,
  Phys.\ Rev.\ D {\bf 73}, 024021 (2006).

\bibitem{Weissenborn:2011ut} 
  S.~Weissenborn, D.~Chatterjee and J.~Schaffner-Bielich,
  Phys.\ Rev.\ C {\bf 85}, 065802 (2012)
  Erratum: [Phys.\ Rev.\ C {\bf 90}, 019904 (2014)].

\bibitem{Alford:2004pf} 
  M.~Alford, M.~Braby, M.~W.~Paris and S.~Reddy,
  Astrophys.\ J.\  {\bf 629}, 969 (2005).
  
\bibitem{Weissenborn:2011qu} 
  S.~Weissenborn, I.~Sagert, G.~Pagliara, M.~Hempel and J.~Schaffner-Bielich,
  Astrophys.\ J.\  {\bf 740}, L14 (2011).

\bibitem{Paschalidis:2017qmb} 
  V.~Paschalidis, K.~Yagi, D.~Alvarez-Castillo, D.~B.~Blaschke and A.~Sedrakian,
  Phys.\ Rev.\ D {\bf 97}, 084038 (2018).

\bibitem{Demorest:2010bx} 
  P.~Demorest, T.~Pennucci, S.~Ransom, M.~Roberts and J.~Hessels,
  Nature {\bf 467}, 1081 (2010).

\bibitem{Antoniadis:2013pzd} 
  J.~Antoniadis {\it et al.},
  Science {\bf 340}, 6131 (2013).

\bibitem{Lattimer:2013hma} 
  J.~M.~Lattimer and A.~W.~Steiner,
  Astrophys.\ J.\  {\bf 784}, 123 (2014).

\bibitem{Ozel:2016oaf} 
  F.~\"Ozel and P.~Freire,
  Ann.\ Rev.\ Astron.\ Astrophys.\ {\bf 54}, 401 (2016).

\bibitem{Most:2018hfd} 
  E.~R.~Most, L.~R.~Weih, L.~Rezzolla and J.~Schaffner-Bielich,
  arXiv:1803.00549 [gr-qc].
  
\bibitem{Radice:2017lry} 
  D.~Radice, A.~Perego, F.~Zappa and S.~Bernuzzi,
  Astrophys.\ J.\  {\bf 852}, L29 (2018)

\bibitem{Tews:2018mr} 
  I.~Tews, J.~Marguenron, and S.~Reddy,
  [arXiv:1804.02783v1 [nucl-th]].

\bibitem{Annala:2018gkv} 
  E.~Annala, T.~Gorda, A.~Kurkela, and A.~Vuorinen,
  Phys.\ Rev.\ D {\bf 120}, 172703 (2018).

\bibitem{Coughlin:2018etal} 
  M. W.~Coughlin \textit{et} al.,
 [arXiv:1805.09371v1 [astro-ph.HE]].

\bibitem{TheLIGOScientific:2018eos1} 
  B.~P.~Abbott {\it et al.} 
  [ arXiv:1805.11581 [gr-qc]].

\bibitem{Ascher:1979iha} 
  U.~Ascher, J.~Christiansen and R.~D.~Russell,
  Math.\ Comput.\  {\bf 33}, no. 146, 659 (1979).
  
  \bibitem{Read:2008iy} 
  J.~S.~Read, B.~D.~Lackey, B.~J.~Owen and J.~L.~Friedman,
  Phys.\ Rev.\ D {\bf 79}, 124032 (2009).

\bibitem{Lattimer:2004nj}
  J.~M.~Lattimer and B.~F.~Schutz,
  Astrophys.\ J.\  {\bf 629}, 979 (2005).

\bibitem{Breu:2016ufb}
  C.~Breu and L.~Rezzolla,
  Mon.\ Not.\ Roy.\ Astron.\ Soc.\  {\bf 459}, 646 (2016).


\bibitem{Lau:2009bu} 
  H.~K.~Lau, P.~T.~Leung and L.~M.~Lin,
  Astrophys.\ J.\  {\bf 714}, 1234 (2010)


\end{thebibliography}
\end{document}